\documentclass[12pt]{article}\usepackage{graphicx,amsmath,amssymb,cite}
\usepackage{epsfig,epsf}
\usepackage{graphicx}
\usepackage{epstopdf}
\newcommand{\beq}{\begin{equation}}
\newcommand{\eeq}{\end{equation}}
\newcommand{\alphabar}{\bar{\alpha}_s}
\newcommand{\betabar}{\bar{\beta_0}}

\numberwithin{equation}{section}

\begin{document}

\vspace*{0.5 cm}

\begin{center}

{\Large{\bf The Effect of the Infrared Phase of the Discrete BFKL Pomeron
 on Transverse Momentum Diffusion}}

\vspace*{1 cm}

{\large Douglas~A.~Ross~$^1$ and Agust{\'\i}n~Sabio~Vera~$^2$} \\ [0.5cm]
{\it $^1$ School of Physics \& Astronomy, University of Southampton,\\Highfield, Southampton SO17 1BJ, UK}\\[0.1cm]
{\it $^2$ Instituto de F{\' \i}sica Te\'{o}rica UAM/CSIC, Nicol{\' a}s Cabrera 15\\ \& Universidad Aut{\' o}noma de Madrid, E-28049 Madrid, ~Spain}\\[0.1cm]
 \end{center}

\vspace*{2 cm}

\begin{abstract}
Imposing infrared boundary conditions on the BFKL equation with running coupling transforms the complex momentum $\omega$-plane cut present in the gluon Green function into an infinite series of positive Regge poles. In addition, a cut on the negative 
$\omega$ line remains. We consider a Hermitian kernel at leading order with running coupling and construct the gluon Green function performing the $\omega$ integration away from the real axis. We find a strong dependence of the asymptotic intercepts and collinear behaviour on the non-perturbative choice of the boundary conditions, in the form of an infrared phase. This is particularly manifest in the asymmetric infrared/ultraviolet structure of the associated diffusion in transverse momentum. We find that random walks into the infrared region are largely reduced in this approach.  
 \end{abstract}


\vspace*{3 cm}

\begin{flushleft}
  May 2016 \\ 
\end{flushleft}

\newpage

\section{Introduction}

A very active area of research in QCD at high energies is the study of diffractive scattering and small $x$ parton distribution functions. Both can be described using the so-called BFKL formalism~\cite{BFKL}. This resummation program can be also applied to the description of high-multiplicity processes such as the inclusive production of Mueller-Navelet jets at hadron colliders~\cite{N.Cartiglia:2015gve}. The study of all of this rich phenomenology at the LHC has opened up a window of opportunity to solve many interesting questions relevant in the perturbative Regge limit of QCD. A pressing one is to find the best treatment of the running of the coupling in this class of processes characterized by the presence of multiple relevant scales. A challenging proposal was put forward in a series of papers where the universal BFKL gluon Green function is modified in the infrared to allow for the existence of Regge poles instead of a dominant cut in the complex momentum plane. In order to shed more light on this approach in this work we investigate 
the associated diffusion into infrared and ultraviolet scales for the typical transverse momentum carried by the $t$-channel reggeized gluons. Of particular importance is to find out how the chosen boundary conditions, expressed in the form of a non-perturbative phase that we will call $\eta$ below, affect the growth with energy and collinear behaviour of the solution to the BFKL equation. Recent interesting studies investigating the possible origin of this non-perturbative phase introducing massive gluons can be found in Ref.~\cite{Levin:2014bwa}.

Let us first introduce some basic ideas on the diffusion properties of the BFKL equation making use of the leading order kernel with fixed coupling. In this case the rapidity dependence of the forward BFKL amplitude ${\cal G}(Y,t_1,t_2)$, which corresponds to the scattering amplitude  for a gluon
 with transverse  momentum $k_1$
 and a gluon with transverse  momentum $k_2$, separated by a rapidity difference $Y$,  obeys the equation
\beq \frac{\partial}{\partial Y} {\cal G}(Y,t_1,t_2) 
\ = \ 
 \alphabar \int dt \,  {\cal K}(t_1,t)   \, 
{\cal G}(Y,t,t_2)  
\eeq
where we have introduced the notation
 $t_i \equiv \ln(k_i^2/\Lambda^2_{\mathrm{QCD}})$, which we use henceforth.
 $ \alphabar \ \equiv \  \frac{C_A}{\pi}  \alpha_s, $
and the kernel ${\cal K}$  has eigenfunctions $e^{i\nu t}$
with eigenvalues
$$\chi(\nu) \ = \ 2 \Psi(1)-\Psi\left(\frac{1}{2}+i\nu\right)
 -\Psi\left(\frac{1}{2}-i\nu\right). $$
For transparency, we will, in the following, work under the ``diffusion'' approximation
in which the expansion of the characteristic function $\chi(\nu)$ is truncated
at quadratic order
$$ \chi(\nu) \ \approx 4\ln 2 - 14 \zeta(3) \nu^2. $$
A  more technically involved calculation using the full characteristic function will be presented in
a future publication. In the quadratic approximation the Green function may be expressed analytically
as
\beq {\cal G} (Y,t_1,t_2) \ = \
\frac{e^{4\ln 2  \, \alphabar  Y} }{\sqrt{56 \zeta(3) \pi  \alphabar Y} }
 \exp  
 \left\{
 \frac{-(t_1-t_2)^2}{56\zeta (3) \alphabar Y  }
 \right\}.
 \eeq

\begin{figure}
\begin{center}
\includegraphics[width=10.0cm]{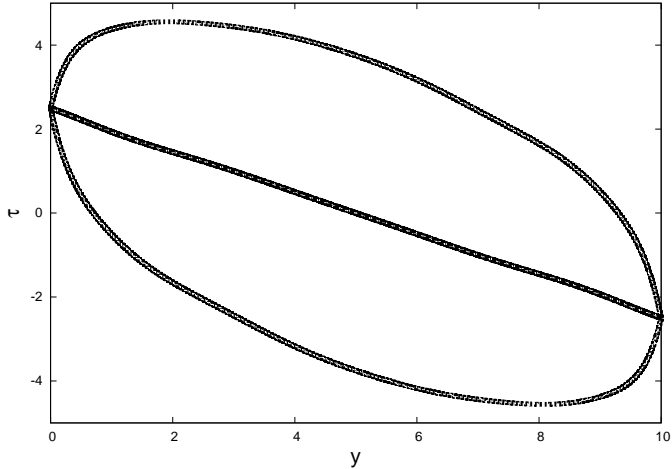}  
\caption{Example of Bartels cigar for fixed coupling with rapidity span $Y=10$. $\tau$ measures the separation from the incoming momenta $t_1$ and $t_2$. The plot is normalized to have a central mean value of $\tau =0$ at  $y=5$.}
\end{center} 
\label{CigarLO}
\end{figure}
Using the Green function property
\beq {\cal G}\left(Y,t_1,t_2\right) \ = \ 
  \int dt \,  {\cal G}\left(y, t_1,t\right) {\cal G}\left(Y-y,t,t_2\right) \label{green_prop}, \eeq
which is valid for any value of rapidity $y$ between $y=0$ and $y=Y$, we see, by examining the integrand on the RHS of (\ref{green_prop})
that for a rapidity $y$, the amplitude has its maximal contribution for transverse momentum given by
\beq \overline{t}(y) \ = \ t_1 + y \frac{\left(t_2-t_1\right)}{Y}, \eeq
and a width, $\Gamma$, where
\beq \Gamma(y) = \sqrt{56 \zeta (3) y (Y-y)/Y}.\eeq
An example of such a distribution is shown in Fig.~{\ref{CigarLO} and is known as ``Bartel's cigar''~\cite{bartels}.
We note that the width vanishes at the ends of the ``cigar'', {\it i.e.}, for $y=0$ or $y=Y$. In the case
 of fixed coupling, the distribution in $t$ is symmetric, {\it i.e.} for any $y$ we have a Gaussian distribution
with width $\Gamma(y)$. 

In the next Section we will introduce the Discrete BFKL pomeron approach and explain how it can be studied using 
integration over contours living away from the real axis. 

\section{The Discrete BFKL pomeron in the complex $\omega$ plane}

If we take into account the running of the coupling using a hermitian kernel, the Green function equation becomes
\beq \frac{\partial}{\partial Y} {\cal G}(Y,t_1,t_2) 
\ = \ 
 \frac{1}{\sqrt{\betabar t_1}}
  \int dt \, {\cal K}(t_1,t)   \frac{1}{\sqrt{\betabar t}}    \, 
{\cal G}(Y,t,t_2) 
\eeq
where we have neglected all heavy fermion threshold effects and written the running coupling, to leading order, as
 $ \alphabar(t) \ = \ \frac{1}{\betabar t} $.

As has been shown in refs.\cite{KLR1} and \cite{KLR2},  the Mellin transform of
this Green function, ${\cal G}_\omega(t_1,t_2)$ can be expressed in terms of 
 Airy functions
\beq {\cal G}_\omega\left( t_1,t_2\right) \ = \ 
 \frac{\pi}{4} 
\frac{\sqrt{t_1t_2}}{\omega^{1/3}} 
\left( \frac{\betabar }{14 \zeta(3)} \right)^{2/3} 
 Ai\left(z(t_1) \right) Bi\left(z(t_2) \right)  \theta\left(t_1-t_2\right)
 \, + \, t_1 \leftrightarrow t_2  \label{mellin-green-1}\eeq
with
 $$ z(t) \ \equiv \left(\frac{\betabar \omega}{14\zeta(3) }\right)^{1/3}
 \left(t-\frac{4\ln 2}{\betabar \omega} \right). $$
As pointed out in \cite{CCS}, with the running coupling the distribution in transverse momentum
is no longer symmetric owing to the fact that there is an enhancement of the amplitude for smaller
 $t$ due to the increase in the running of the coupling. The symmetric ``cigar'' shape becomes
 more like a ``banana" and for sufficiently small values of $t_1$ and $t_2$ the shape can
 tunnel into an entirely new shape.

It was pointed out in \cite{KLR1} and \cite{KLR2} that whereas  the Mellin transform of the Green function
 given by (\ref{mellin-green-1}) satisfies the required UV boundary condition that it vanishes as $t_1\to\infty$
 or $t_2\to\infty$, it is not unique. Indeed one can add to it any solution of the homogeneous part of the Green function  
equation with the same ultraviolet behaviour, {\it i.e.} we can replace the Airy function  $Bi(z)$ by
$$ \overline{Bi}(z) \ \equiv Bi(z) + c(\omega) Ai(z). $$ 
Taking the ($\omega-$dependent) coefficient function $c(\omega)$ to be of the form
\beq c(\omega) \ = \ \cot\left(\eta- \frac{2}{3} \sqrt{\frac{\betabar \omega}{14\zeta(3)}} \left(
 \frac{4\ln 2}{\betabar\omega}-t_0\right)^{3/2}  \right),  
\label{phase} 
 \eeq
introduces a set of discrete poles for the BFKL pomeron for values of $\omega$
 such that
\beq 
\eta- \frac{2}{3} \sqrt{\frac{\betabar \omega}{14\zeta(3)}} \left(
 \frac{4\ln 2}{\betabar\omega}-t_0  \right)^{3/2}  \ =  - n \pi, 
 \label{NPphase}
 \eeq
and at the same time sets the phase at some infrared fixed point $t_0$ in the transverse
 momentum to $\frac{\pi}{4}+\eta$. As pointed out in ref.\cite{lipatov86}, the value of 
 $\eta$ is determined by the non-perturbative properties of QCD. It cannot be calculated
 {\it a priori} in perturbative QCD and has to be left as a free parameter in a fit to data.
\begin{figure}
\begin{center}
\includegraphics[width=12.0cm]{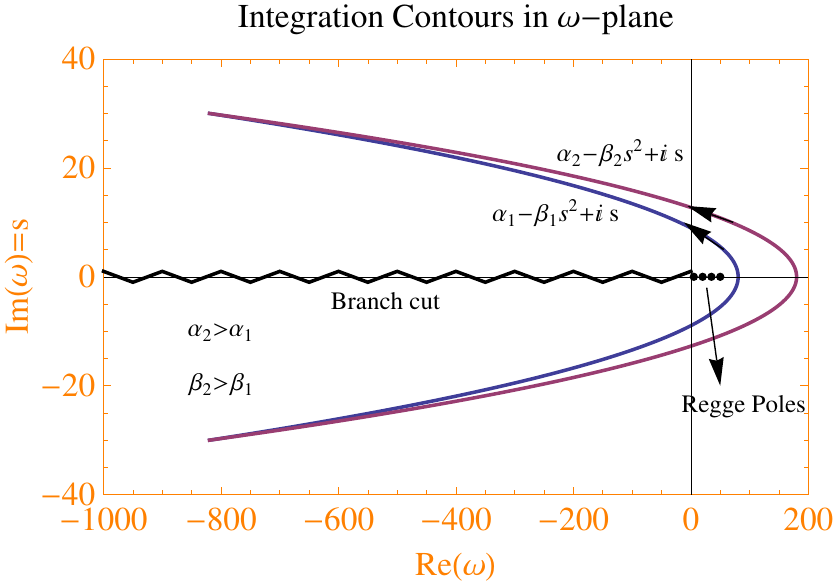}  
\end{center} 
\caption{Integration contours chosen to surround the branch cut and discrete poles. }
\label{Contours}
\end{figure}

As well as the above-mentioned poles, the 
 Mellin transform of the Green function (as a complex function of complex $\omega$)
has a discontinuity across the negative real axis whenever $z(t_1)$ or $z(t_2)$ is negative.
 In order to ensure that this cut and all the poles are correctly accounted for
we invert the Mellin transform by performing the integral (numerically)
 \beq {\cal G}(Y,t_1,t_2) \ = \ \frac{1}{2\pi i} \int_{\cal C} d\omega \, e^{\omega Y} {\cal G}_\omega(t_1,t_2) \eeq
 where ${\cal C}$ is a contour that intercepts the real axis to the right of all poles and any
branch-point and surrounds the negative real axis as shown in Fig.~\ref{Contours}.
 We have chosen
 \beq \omega \ =  \frac{1}{2}  - \mu^2 + i \mu, \ \ -\infty \, < \, \mu \, < \infty. \eeq
\begin{figure}
\begin{center}
\includegraphics[width=12.0cm]{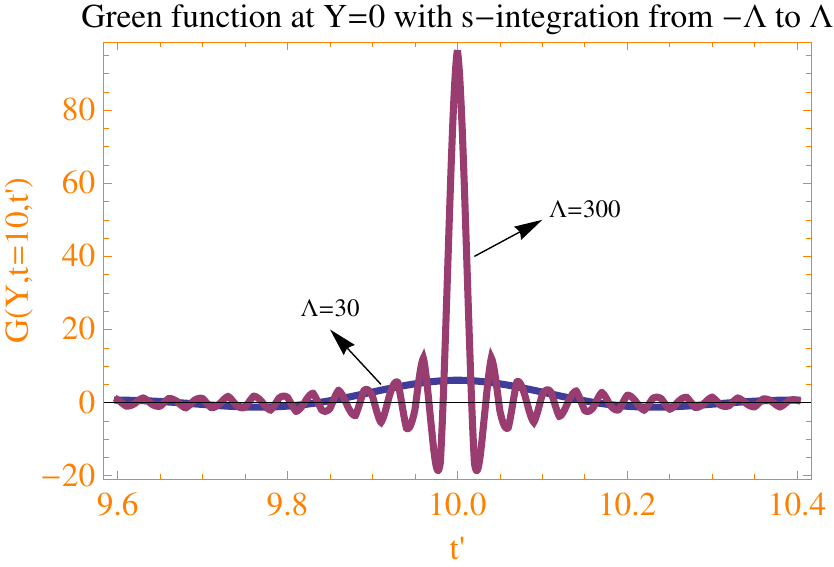}  
\end{center} 
\caption{ Two approximations to the $\delta$-function for $Y=0$ integrating in the region $- \Lambda < \mu < \Lambda$ with $\Lambda=30, 300$.  }
\label{DeltafunctionLambda}
\end{figure}
We have conducted two  checks of our numerical integration. Firstly, we have
 varied the intercept and the slope of the contour to ensure no change (these are represented by $\alpha_i$ and 
 $\beta_i$ in Fig.~\ref{Contours}). Secondly, we
have reproduced to a good approximation the expected $\delta(t_1-t_2)$  behaviour in the case $Y=0$.
 An exact $\delta-$function is really only obtained if one integrates along an infinite contour.
 In Fig. ~\ref{DeltafunctionLambda}, we show the approximation we obtain when $|\mu|_{\mathrm{max}}=\Lambda$ is taken
 to be 300 (corresponding to a minimum value of $\Re e \{\omega\}$ of $- 10^5$) and we compare it with a regularized $\delta$-function for $\Lambda=30$ (corresponding to a minimum value of $\Re e \{\omega\}$ of $- 900$). We see that 
in order to obtain a reasonable approximation to the expected $\delta$-function we need to integrate
 over a very large portion of the contour. 
 \begin{figure}
\begin{center}
\includegraphics[width=12.0cm]{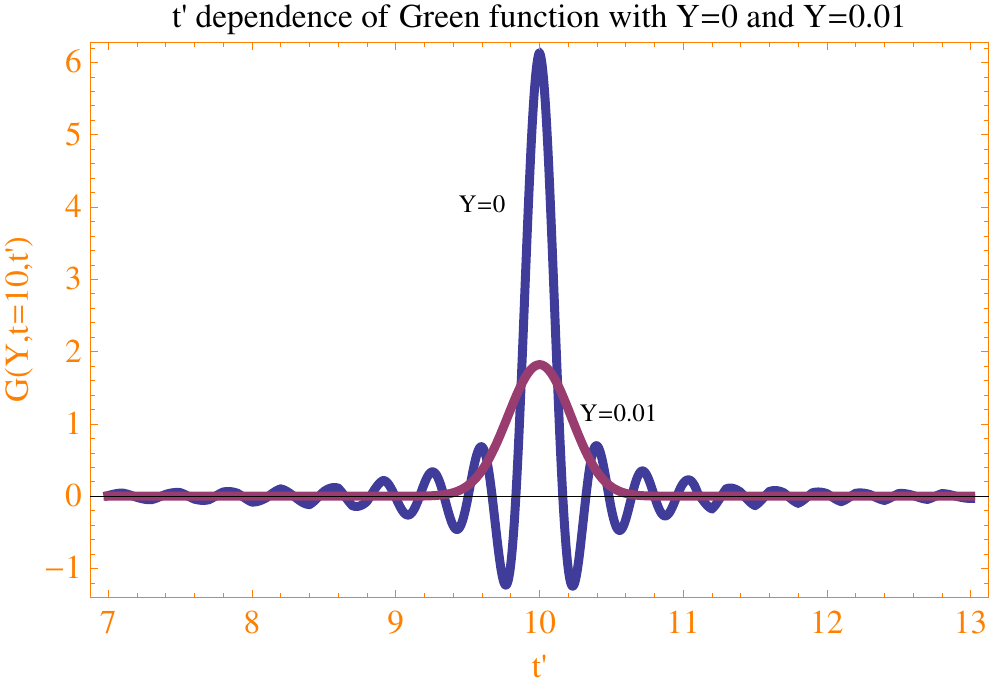}  
\end{center} 
\caption{Sudden spread of the $\delta-$function when we vary $Y$ from 0 to 0.01.}
\label{GGFDelta}
\end{figure}
However, for very small but non-zero values of rapidity $Y$, the integrand converges quite quickly and the $\delta$-function becomes very wide. We have illustrated this point in Fig.~\ref{GGFDelta}. 

\subsection{The non-perturbative phase and its effect on the Green function}

As we have explained above the non-perturbative parameter $\eta$ affects the phase in Eq.~(\ref{NPphase}) and has an important influence on the structure of the gluon Green function. 
\begin{figure}
\begin{center}
\includegraphics[width=12.0cm]{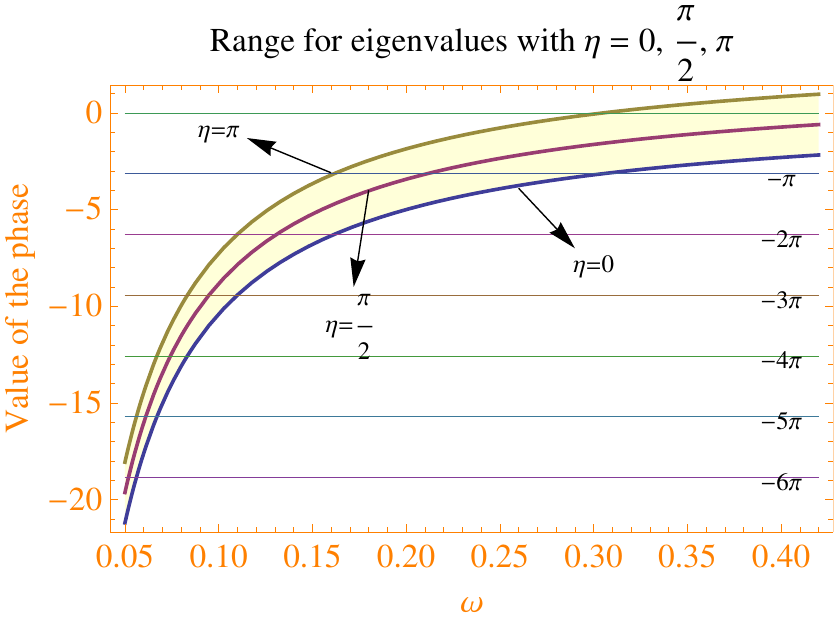}  
\end{center} 
\caption{The spectrum of eigenvalues corresponds to the intersection of the phase in Eq.~(\ref{NPphase}) with the lines $-n \pi$ for positive integer $n$. $\omega_1-\omega_2$ is largely reduced when moving from $\eta=0$ to $\eta=\pi$.}
\label{EigenvaluesEta}
\end{figure}
 The most important consequence is that the solution for $\omega$ in Eq.~(\ref{NPphase}), which corresponds to a discrete spectrum of positive $\omega_n$, gradually reduces the gap between the first eigenvalue, $\omega_1$ 
(which dominates at
 sufficently  large $Y$) since it is the most positive) and the second eigenvalue $\omega_2$ as we increase $\eta$ from 0 to $\pi$. We show the $\eta$ dependence of this spectrum of eigenvalues in Fig.~\ref{EigenvaluesEta}. 
 \begin{figure}
\begin{center}
\includegraphics[width=12.0cm]{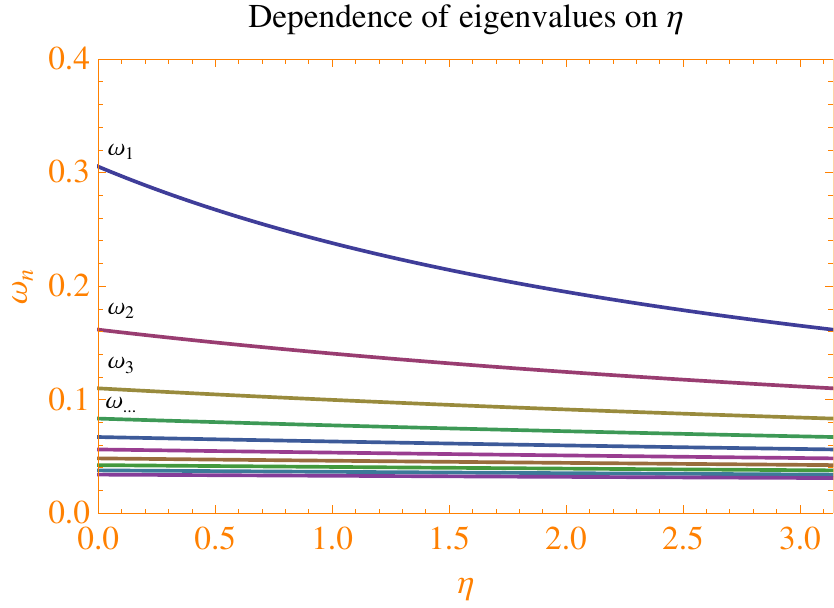}  
\end{center} 
\caption{$\eta$ dependence in the spectrum of eigenvalues.}
\label{WnvsEta}
\end{figure}
 To further clarify this important point we show the $\eta$ dependence of the first eigenvalues in Fig.~\ref{WnvsEta}.

In terms of the gluon Green function it is clear that increasing the value of $\eta$ reduces the asymptotic rise with energy. We show this for $t=10,t'=8$ in Fig.~\ref{GGFvsY}.
\begin{figure}
\begin{center}
\includegraphics[width=12.0cm]{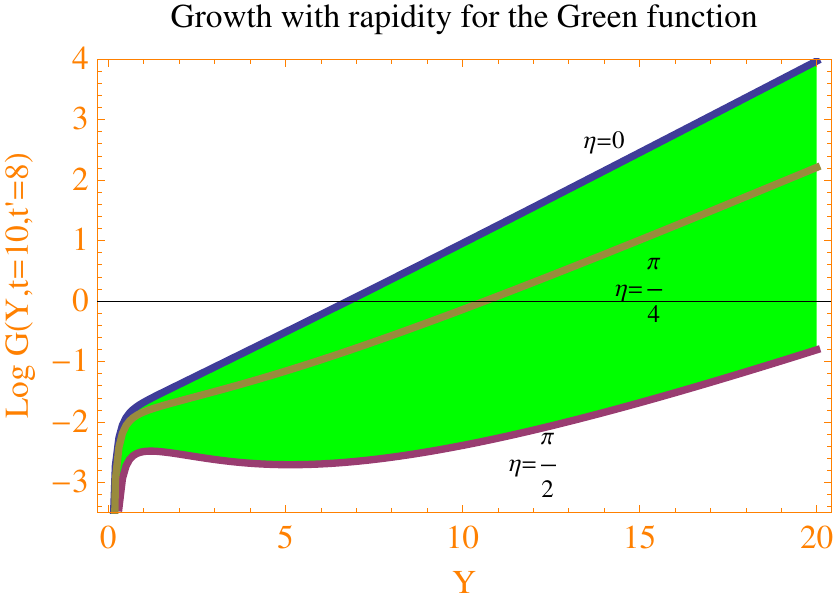}  
\end{center} 
\caption{Growth with energy of the solution to the BFKL equation with running coupling in the Discrete pomeron approach.}
\label{GGFvsY}
\end{figure}
There is a range of allowed growths  in-between the two lines which can only be fixed by a phenomenological comparison to data. 

In order to fix the position of the integration contours, in particular the point $\omega_0$ where they cross the real axis (see Fig.~\ref{Contours}), we should note that Eq.~(\ref{NPphase}) is an approximation of the corresponding exact relation
\begin{eqnarray}
\eta - \int^{t_c}_{t_0} \nu_\omega (t')  dt'  &=& - n \pi,
\end{eqnarray}
where 
\begin{eqnarray}
  \chi(\nu_\omega (t)) &=& \omega {\bar{\beta_0} } t ~\simeq~ \omega {\bar{\beta_0} } t_c+ \frac{\chi''(0)}{2} \nu_\omega^2 (t), 
\end{eqnarray}
which is valid within the  quadratic approximation in the BFKL kernel. In more detail,  since 
\begin{eqnarray}
\nu_\omega (t) \simeq \sqrt{\frac{2}{\chi''(0)}} \sqrt{ \omega {\bar{\beta_0} }(t-t_c)}, 
\end{eqnarray}
then 
 \begin{eqnarray}
\eta - \int_{t_0}^{t_c} \nu_\omega (t')  dt'  &\simeq&  \eta - \frac{2}{3} \sqrt{\frac{\omega {\bar{\beta_0} }}{14 \zeta(3)}} \left(t_c-t_0 \right)^{\frac{3}{2}}.
\end{eqnarray}
As we have recently discussed in Ref.~\cite{Ross:2016kzz}, the value $t_c= 4 \ln{2}/\bar{\beta_0} \omega$ is the turning point (in the $t$ variable) from an oscillatory behaviour of the eigenfunctions to a decaying one. To allow for a correct transition from the non-perturbative region around $t_0$ to a perturbative one above $t_c$ we choose contours to the right of all the poles with $n>0$ in Eq.~(\ref{NPphase}) and to the left of the pole associated with $n=0$ (we cross the real axis at a point $\omega_0 < 4 \ln{2}/\bar{\beta_0} t_0$), ensuring, in this way, that $t_c>t_0$. This pole for $n=0$ only appears when $\eta$ is larger
 than some critical value, $\eta_c$.

The collinear behaviour of the Green function is investigated in Fig.~\ref{Alcala2016DouglasCigarGGFvst}. 
$t$ has been fixed to 10 and $t'$ varies from small to large values for $Y=10$. Note that the effect of $\eta$ is much more relevant at small values of $t$, which is expected, since the value of $\eta$ encodes the infrared behaviour. The Green function suffers a strong reduction as we increase 
$\eta$, reaching the point of being negative at small values of $t$ when $\eta$ gets close to $\pi/2$. 
\begin{figure}
\begin{center}
\includegraphics[width=12.0cm]{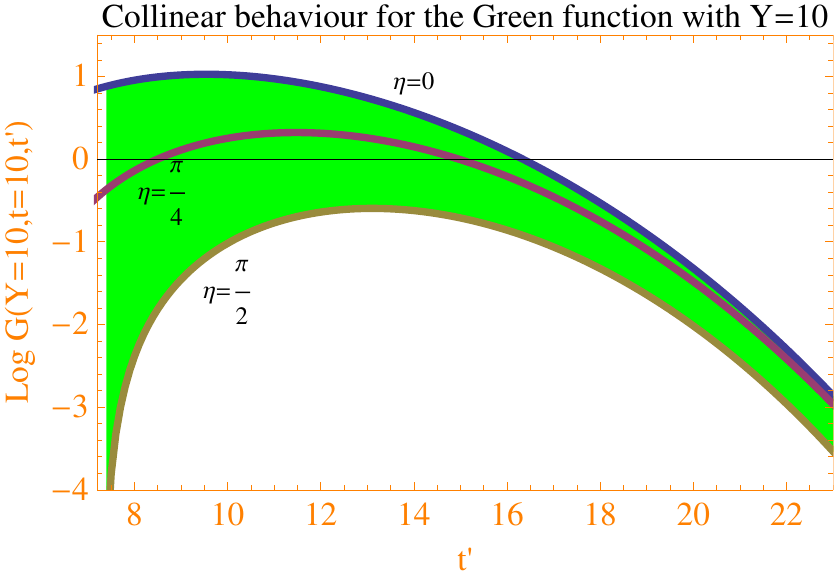}  
\end{center} 
\caption{Collinear behaviour of the solution to the BFKL equation with running coupling in the Discrete pomeron approach.}
\label{Alcala2016DouglasCigarGGFvst}
\end{figure}
This indicates that only small positive values of $\eta$ are allowed if physical cross sections are a consequence of our analysis. 

\section{Numerical results: Diffusion distributions}

The $t$-profile of our Green function contains some important information about the discrete pomeron approach:  it permits the  construction of  the diffusion profile of our solutions. For some given values of  external transverse scales $t$ and $t'$, this corresponds to the extent to which the internal propagators in the Green function deviate from a mean value when we probe the quantity at an internal rapidity $0 \leq y \leq Y$, ({\it i.e.} smaller than the total rapidity). The way to proceed is to make use of the integrand of Eq.~(\ref{green_prop}) and identify its maximum. Then we calculate the two transverse momentum values to the left and right of this point where the Green function equals
 its maximum value divided by $e$. We define the difference between these values of the transverse
momenta and the mean value to be the half-width. Importantly, the distribution is not, in general, symmetric, so that the ultraviolet and infrared half-widths are unequal.
\begin{figure}
\begin{center}
\includegraphics[width=10.0cm]{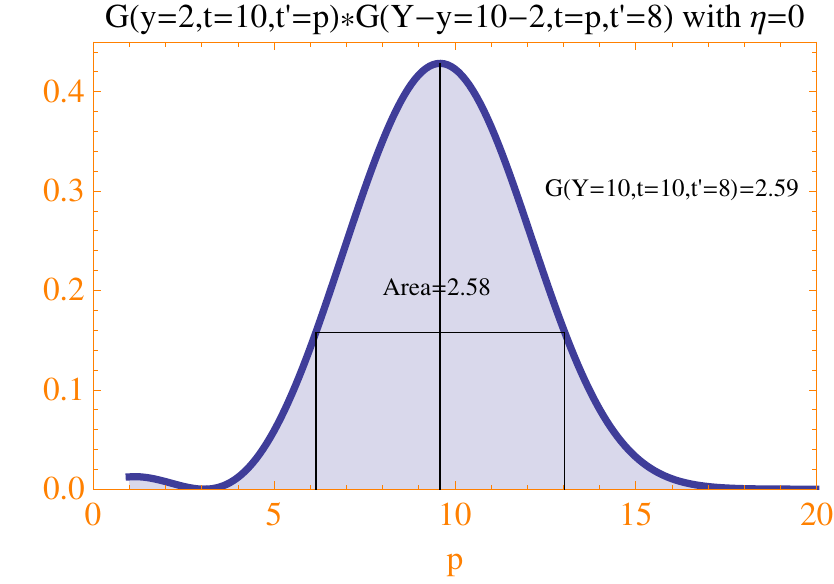}\\
\includegraphics[width=10.0cm]{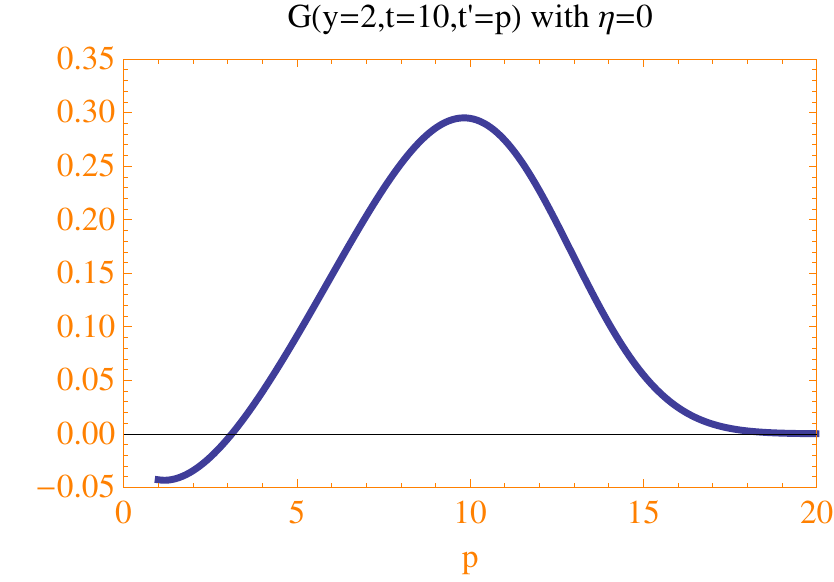} \includegraphics[width=10.0cm]{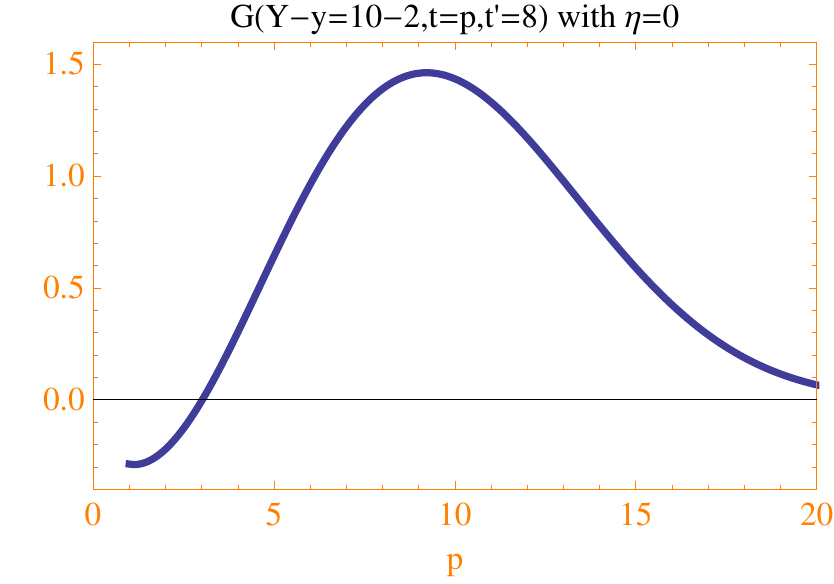}   
\end{center} 
\caption{Graphical example to show the consistency with Eq.~(\ref{green_prop}) in the construcion of diffusion plots. }
\label{CigarGGFArea}
\end{figure}

We have  shown this procedure graphically in the example of Fig.~\ref{CigarGGFArea}. The middle and bottom plots correspond to the $p$-profile of the two Green functions evaluated at different values $y$ and $Y-y$ of rapidity. The product
of these profiles  generates the needed integrand (in $t$)
 shown in the   top plot of the Figure. By virtue of Eq.~(\ref{green_prop}), the area under the curve on the top plot
  returns the value of the Green function at rapidity $Y$
 with the two external transverse momenta set to the fixed transverse momenta of the lower two plots.
 Note that the chosen external scales are $t=10$ and $t'=8$ and we find that the maximum of the integrand lies between those two values. A crucial test of the validity of our results is that the integration over $p$ of the top plot should coincide with the value of the Green function at $Y=10$, independently of the value of $y$ (in this case we have chosen $y=2$ but the final results do not depend on this choice). For this particular example we obtain an area of 2.58 (corresponding to the integration in Eq.~(\ref{green_prop})) whereas the directly calculated
value for the Green function  for $t=10$ and $t'=8$   of 2.59. This indicates the level of accuracy of our numerical methods. 

This procedure can be iterated for different values of $t$ and $t'$ and a selection of values for $\eta$. This is what
 has been done in Fig.~\ref{CigarsPositiveEta}.
\begin{figure}
\vspace{-2cm}
\begin{center}
\includegraphics[width=7.0cm]{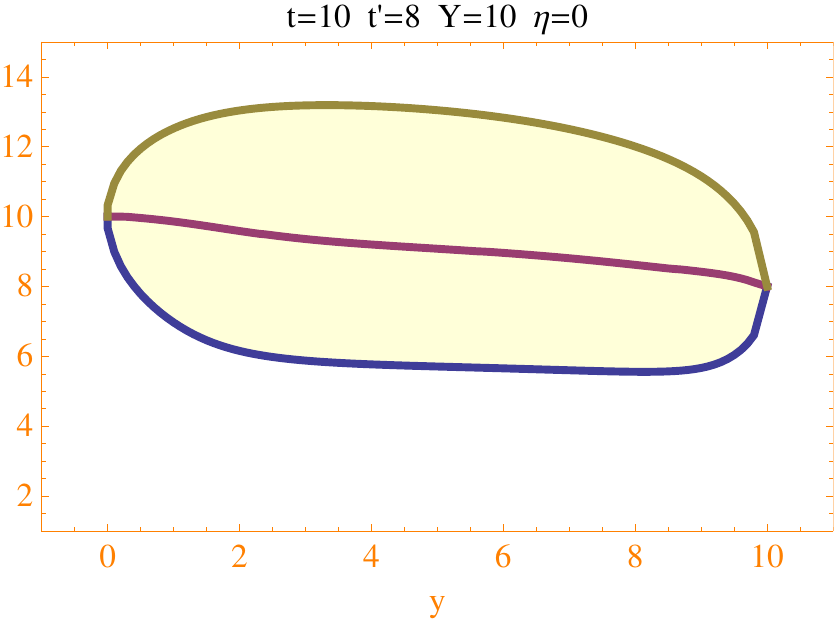}  \includegraphics[width=7.0cm]{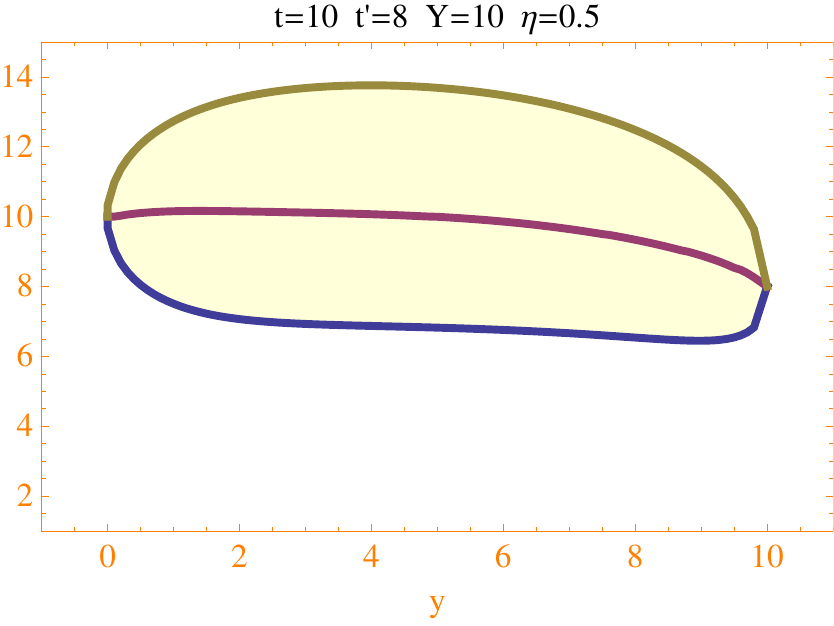} \\
\includegraphics[width=7.0cm]{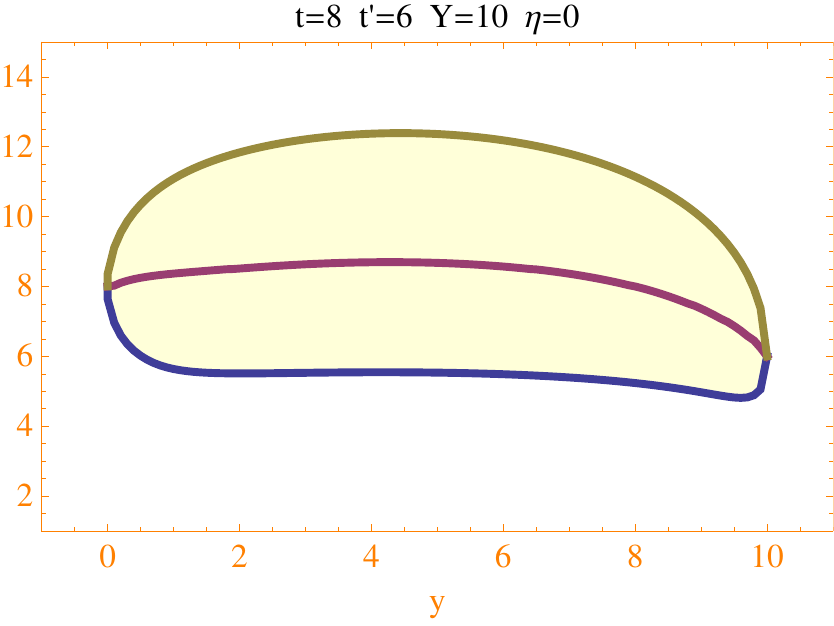} \includegraphics[width=7.0cm]{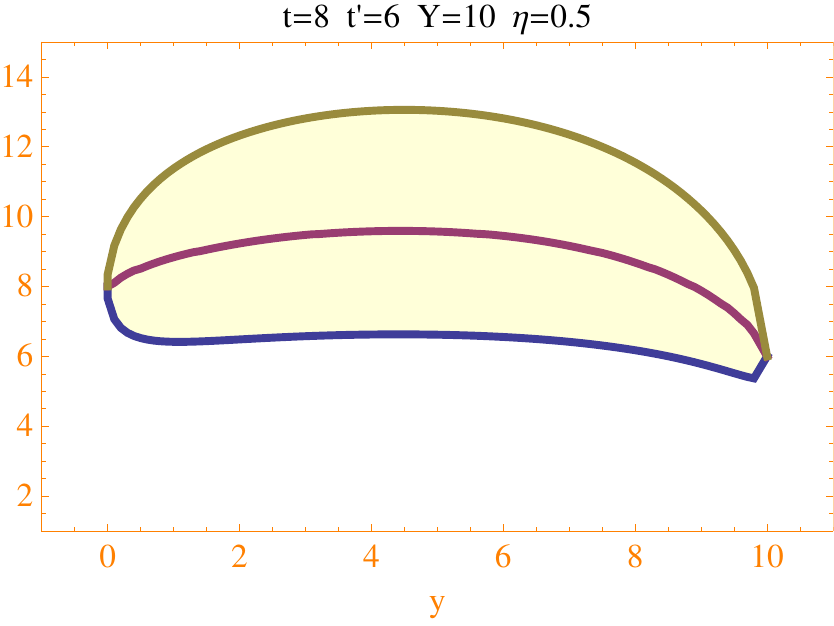}   \\
\includegraphics[width=7.0cm]{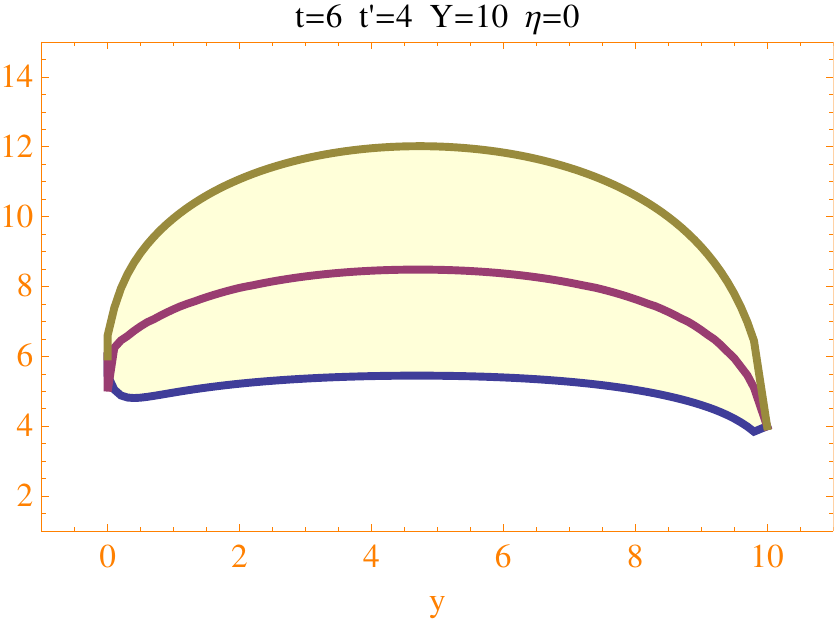} \includegraphics[width=7.0cm]{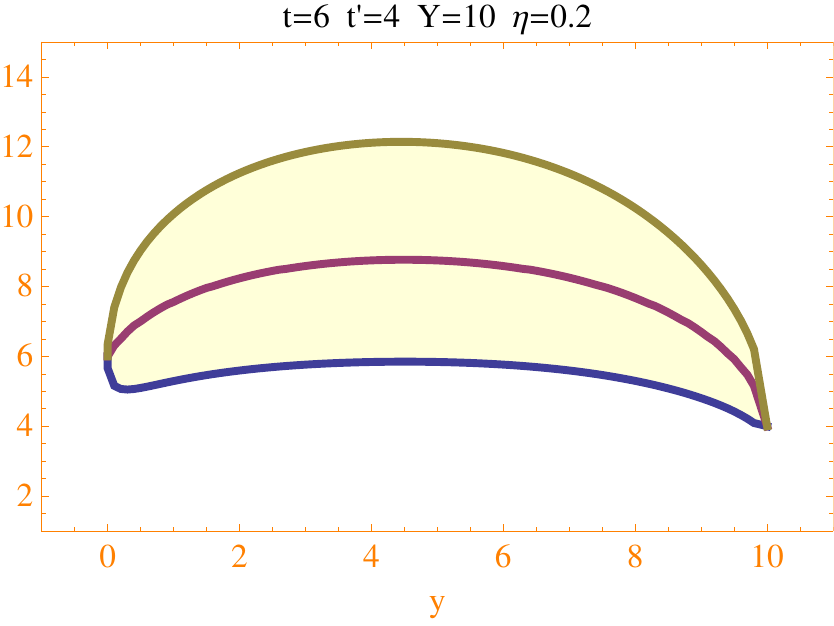}   \\
\includegraphics[width=7.0cm]{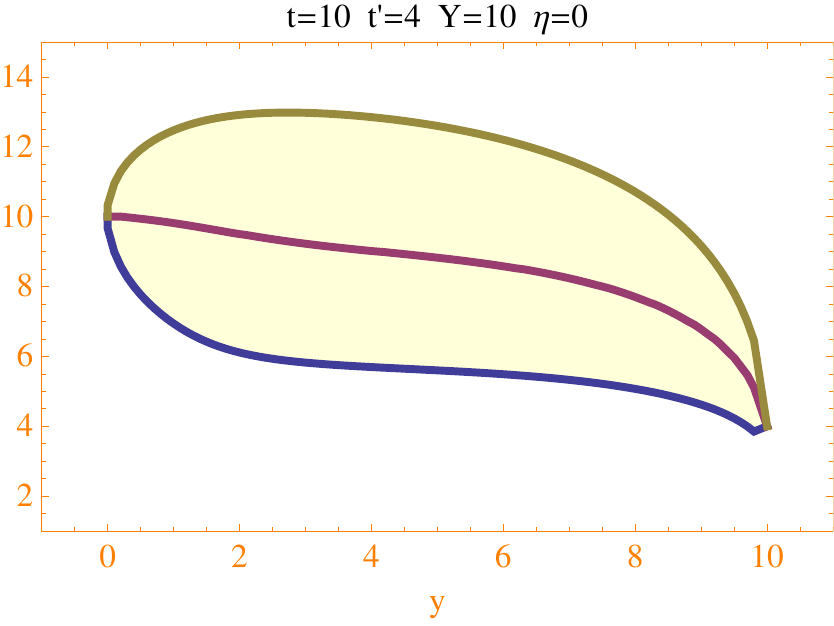} \includegraphics[width=7.0cm]{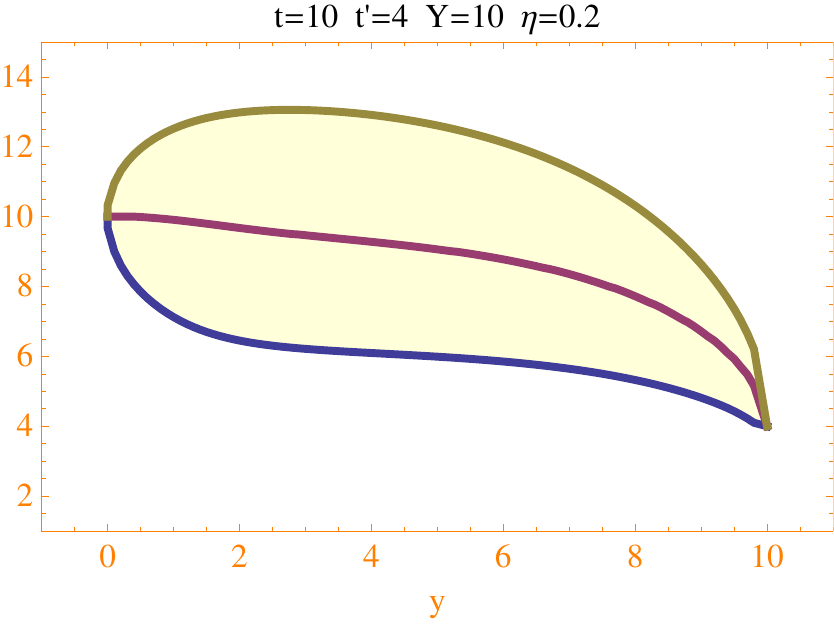}    
\end{center} 
\caption{Diffusion figures for different values of $t$ and $t'$. From top to bottom we lower these external scales. The LHS plots correspond to a non-perturbative phase $\eta=0$ and the RHS ones to $\eta=0.2, 0.5$. The two plots at the bottom correspond to a more Deep Inelastic (DIS)-like situation with $t=10$ and $t'=4$.}
\label{CigarsPositiveEta}
\end{figure}
This plot contains abundant information about the physics introduced by the discrete pomeron approach. We observe that when we have large external scales (top plots in the figure) we have a rather ultraviolet/infrared symmetric diffusion profile. However we start to notice a small suppression of the diffusion towards lower scales as it can be deduced from the flatter shape in that region for both $\eta=0, 0.5$ plots. This suppression of the infrared physics is stronger as we move the full BFKL ladder downwards in the characteric scales. This can be seen from the shape of the different plots in Fig.~\ref{CigarsPositiveEta} where even the mean values of the distributions are pushed towards harder scales than the external ones. This effect is slightly more accute in the case with a small positive value of $\eta$.

It is possible to extend our approach to small negative values of $\eta$. We see this in Fig.~\ref{CigarT10S4Y10Etam}
 where we compare the cases $\eta=-0.2$ with $\eta=+0.2$ for two different pairs of external
 tranverse momenta $t$ and $t^\prime$. In both cases the end of the plot with larger values of $t$ are very similar. However, in the regions close to smaller external values of $t'$ the distributions
  show a slightly different, more ``round", shape. We have found that this is due to a smaller negative region at low values of $t$ in the gluon Green function as compared to the positive $\eta$ case. Thinking of $\eta$ as a free parameter to generate physical cross sections, it is possible 
 that the 
negative value is strongly constrained  by the fact that the solution to Eq.~(\ref{NPphase}) generates rather large intercepts 
 in comparison with the positive $\eta$ case. Nevertheless, slightly negative  values are certainly a possibility when scanning the parameter space in future fits to experimental data.
\begin{figure}
\begin{center}
\includegraphics[width=7.0cm]{Alcala2016DouglasCigarT6S4Y10Eta02.pdf} \includegraphics[width=7.0cm]{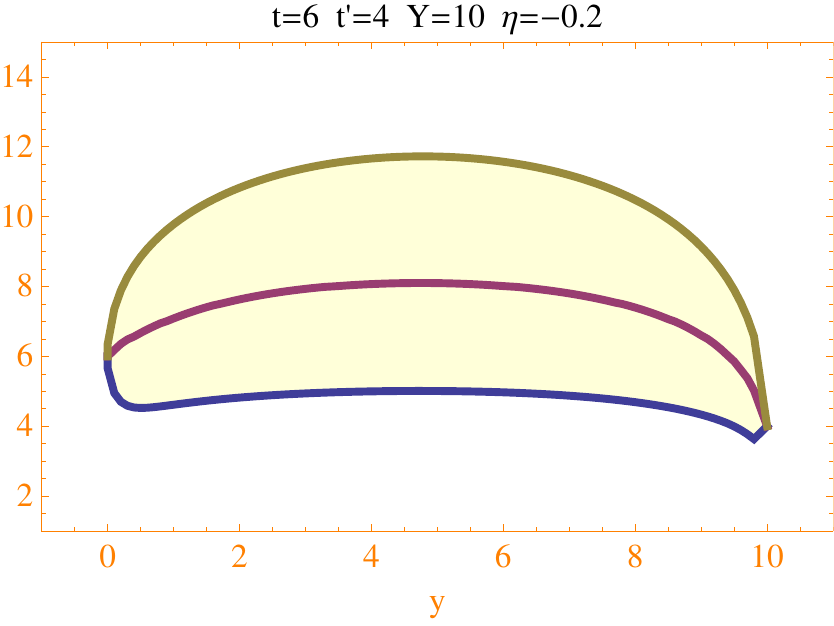}   \\
\includegraphics[width=7.0cm]{Alcala2016DouglasCigarT10S4Y10Eta02.pdf} \includegraphics[width=7.0cm]{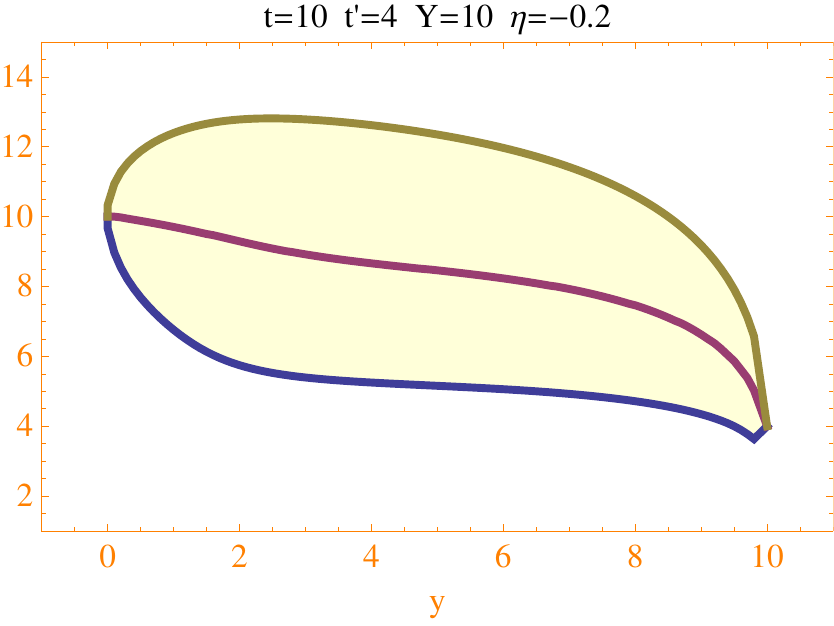}    
\end{center} 
\caption{Diffusion figures for different values of $t$ and $t'$.  The LHS plots correspond to a non-perturbative phase $\eta=0.2$ and the RHS ones to $\eta=-0.2$. }
\label{CigarT10S4Y10Etam}
\end{figure}

\section{Conclusions}

We have performed a detailed analysis of the diffusion properties of the BFKL equation in the 
discrete pomeron approach. This is characterized by the imposition of infrared boundary conditions 
which appear at a fixed scale above the QCD Landau pole. As a consequence, a discretization in the 
BFKL gluon function appears, stemming from the existence of an infinite number of Regge poles in the complex angular momentum $\omega$-plane. However, the analytic structure is more complicated since a branch cut is also present on the negative real axis in the $\omega$-plane. We have taken into account all of these contributions by performing our analysis using contours of integration away from the real axis, which is only crossed at a positive $\omega$ to the right of all the singularities. In this way we do not need to sum over an infinite number of eigenfunctions. In order to simplify our analysis we have worked within a quadratic 
approximation for the BFKL kernel. We have produced results for the behaviour of the gluon Green function as a function of the rapidity $Y$ and transverse momenta. The latter allowed us to show that in the discrete pomeron approach there is an effective barrier which prevents the propagators of the off-shell reggeized gluons to random walk towards very soft regions. This effect is enhanced if we make use of the non-perturbative phase which is present in our approach but we are not able to fix from first principles. It will be very interesting to find the best fit to experimental observables using this free parameter as a new degree of freedom. The rich phenomenology in multi-regge kinematics present at the Large Hadron Collider should help in the pursuit of this program.

\section*{Acknowledgements}
One of us (DAR) wishes to thank the Institute for Theoretical Physics
at the Autonomous University of Madrid for its hospitality during the time
that this work was carried out,  as well as the Leverhulme Trust for an Emeritus Fellowship.  ASV acknowledges support from the Spanish Government (MICINN (FPA2015-65480-P)) and to the Spanish MINECO Centro de Excelencia Severo Ochoa Programme (SEV-2012-0249). We would like to thank the university of Alcal{\'a} de Henares for the use of their facilities at the Colegio de M{\'a}laga.

\end{document}